\documentclass[cha,
 aip,
 amsmath,amssymb,
 reprint,%
floatfix,
]{revtex4-1}
\usepackage{graphicx}% Include figure files
\usepackage{dcolumn}% Align table columns on decimal point
\usepackage{bm}% bold math
\usepackage[mathlines]{lineno}% Enable numbering of text and display math
\usepackage[utf8]{inputenc}
\usepackage[T1]{fontenc}
\DeclareUnicodeCharacter{0394}{$\Delta$}
\usepackage{mathptmx}
\usepackage{etoolbox}
\usepackage[table]{xcolor}
\usepackage{booktabs}
\usepackage{ulem} 
\makeatletter
\def\@email#1#2{%
 \endgroup
 \patchcmd{\titleblock@produce}
  {\frontmatter@RRAPformat}
  {\frontmatter@RRAPformat{\produce@RRAP{*#1\href{mailto:#2}{#2}}}\frontmatter@RRAPformat}
  {}{}
}%
\makeatother
\makeatletter
\let\selectlanguage\@gobble
\makeatother
\begin{document}

\preprint{AIP/123-QED}
%\title{Experimentally Accurate Predictions of Core-Electron Binding Energies in Large Organic Molecules with Equivariant Graph Neural Networks}
%\title{Experimentally Accurate Core-Electron Binding Energy Predictions with an Equivariant Graph Neural Network}
\title{Graph Neural Network Predictions of Carbon 1s Binding Energies with Near-Experimental Accuracy}
\author{Adam E. A. Fouda*}
\affiliation{Department of Physics, The University of Chicago, Chicago, IL 60637, USA}
\affiliation{Chemical Sciences and Engineering Division, Argonne National Laboratory, 9700 S. Cass Avenue, Lemont, IL 60439, USA}
\email{adamfouda@uchicago.edu}
\author{Joshua Zhou}
\affiliation{Pritzker School of Molecular Engineering, The University of Chicago, Chicago, Illinois 60637, United States}
\author{Rodrigo Ferreira}
\affiliation{Chemical Sciences and Engineering Division, Argonne National Laboratory, 9700 S. Cass Avenue, Lemont, IL 60439, USA}
\affiliation{Pritzker School of Molecular Engineering, The University of Chicago, Chicago, Illinois 60637, United States}
\author{Patrick Phillips}
\affiliation{Chemical Sciences and Engineering Division, Argonne National Laboratory, 9700 S. Cass Avenue, Lemont, IL 60439, USA}
\affiliation{Department of Computer Science, The University of Chicago, Chicago, Illinois 60637, United States}
\author{Valay Agarawal}
\affiliation{Department of Chemistry, The University of Chicago, Chicago, Illinois 60637, United States}
\author{Bhavnesh Jangid}
\affiliation{Department of Chemistry, The University of Chicago, Chicago, Illinois 60637, United States}
\author{Jacob J. Wardzala}
\affiliation{Department of Chemistry, The University of Chicago, Chicago, Illinois 60637, United States}
\author{Rui Ding}
\affiliation{Chemical Sciences and Engineering Division, Argonne National Laboratory, 9700 S. Cass Avenue, Lemont, IL 60439, USA}
\affiliation{Pritzker School of Molecular Engineering, The University of Chicago, Chicago, Illinois 60637, United States}
\author{Junhong Chen}
\affiliation{Chemical Sciences and Engineering Division, Argonne National Laboratory, 9700 S. Cass Avenue, Lemont, IL 60439, USA}
\affiliation{Pritzker School of Molecular Engineering, The University of Chicago, Chicago, Illinois 60637, United States}
\author{Nicole Tebaldi}
\affiliation{Data Science Institute, The University of Chicago, Chicago, Illinois 60637, United States}
\author{Phay J. Ho}
\affiliation{Chemical Sciences and Engineering Division, Argonne National Laboratory, 9700 S. Cass Avenue, Lemont, IL 60439, USA}
\author{Laura Gagliardi}
\affiliation{Department of Chemistry, The University of Chicago, Chicago, Illinois 60637, United States}
\affiliation{Pritzker School of Molecular Engineering, The University of Chicago, Chicago, Illinois 60637, United States}
\author{Linda Young}
\affiliation{Chemical Sciences and Engineering Division, Argonne National Laboratory, 9700 S. Cass Avenue, Lemont, IL 60439, USA}
\affiliation{Department of Physics and James Franck Institute, The University of Chicago, Chicago, Illinois 60637, USA}

\date{\today}% It is always \today, today,

\begin{abstract}

Graph neural networks are promising architectures for fast, accurate and transferable predictions of core-electron binding energies, which depend on the local bond environment. Here we present a graph neural network model for predicting carbon 1$s$ core-electron binding energies in organic molecules. The model is trained with multiconfiguration pair-density functional theory on 8637 carbon atoms in 2116 molecules with 4-16 atoms and evaluated against 570 experimental values in 113 different molecules containing 3-45 atoms. Previous work benchmarked a mean absolute error of 0.27 eV to experiment for the training data level of theory [J. Phys. Chem. A 2025, 129, 36, 8419–8431] and the present model demonstrates an experimental evaluation error of 0.33 eV with good size transferability to larger organic molecules. An equivariant graph neural network is benchmarked against its rotationally invariant analogue and a model comprised of the smooth overlap of atomic positions descriptors and kernel ridge regression for training data efficiency and stability to non-equilibrium geometries absent from the training data. All models show good training data efficiency and the graph based models have improved transferability to non-equilibrium geometries. The use of chemically informed, graph-normalized node features reduces the graph neural network's dependence on message-passing depth. A case study on the 45 atom avobenzone tautomers demonstrates the model's ability for instant and precise analysis of complex molecules. The software and data are provided by the open-source AugerNet package at https://doi.org/10.5281/zenodo.19689244.
%Finally, the model's equivariance with respect to 3D rotation is shown to improve predictions on non-equilibrium geometries from the methanol C-O bond stretching surface, compared to a 3D rotation invariant model. 

\end{abstract}

\maketitle

\section{\label{sec:intro}Introduction}

 X-ray photoelectron spectroscopy (XPS) detects the photoelectrons following the ionization of inner-shell orbitals by x-ray irradiation. The widespread use of this characterization technique on materials,\cite{bagus_interpretation_2013} molecules,\cite{siegbahn_esca_1970} solutions,\cite{seidel_valence_2016} liquids,\cite{hurisso_amino_2011,dick_probing_2020} and biological matter\cite{ratner_surface_1983}, is driven by its atom site selectivity which results from the effect of the local bond environment on the core electron binding energy (CEBE). This phenomenon, known as the chemical shift\cite{bagus_mechanisms_1999}, distinguishes XPS peaks of common atom types with different functional groups. However, resolving bond environments at specific atomic sites using XPS alone is challenging and often impeded by overlapping CEBEs for atoms in different environments. The absence of well defined reference data for such cases results from the complex influence of the structural environment on the CEBE which occurs through multiple competing mechanisms, such as charge transfer, electric fields and hybridization\cite{bagus_mechanisms_1999}. 

 Therefore, XPS measurements often rely on computations to support peak assignments. Computational approaches are broadly split in two categories: ``\textit{forward}'' (structure-to-spectrum) and ``\textit{backward}'' (spectrum-to-structure) mapping. ``\textit{Backward}'' mapping extracts chemical information directly from experimental spectra, typically leveraging reference databases and prior knowledge of the investigated system. Recently, this approach has seen a dramatic rise in data driven machine learning (ML) methods, including convolutional neural networks,\cite{pielsticker_convolutional_2023,drera_deep_2020,vakhrushev_application_2024} transformers\cite{simperl_transformer_2025} and large language models\cite{de_curto_large_2024} to establish automated high-throughput workflows for the rapid analysis of XPS spectra.

``\textit{Forward}'' mapping has become a well established approach, particularly in materials and molecular analysis, owing to numerous developments in accurate and efficient \textit{ab-initio} quantum chemistry methods for computing x-ray spectroscopic observables.\cite{rankine_progress_2021} From these techniques, density functional theory (DFT) has gained widespread adoption due to its low cost and demonstrated experimental accuracy.\cite{besley_density_2020,besley_modeling_2021,towers_tompkins_efficient_2025,towers_tompkins_core-level_2025} DFT methods typically achieve experimentally accurate CEBEs when they are employed within the $\Delta$SCF framework,\cite{bagus_self-consistent-field_1965,ljungberg_implementation_2011} which computes the energy difference between the ground and core-ionized state wave functions\cite{fouda_assessment_2017,hanson-heine_basis_2018,hirao_core_2023,hirao_core-level_2023,besley_density_2021,hirao_exploiting_2024-1,jorstad_span_2022,fouda_improving_2020,zheng_benchmark_2022,hirao_theoretical_2025-1}. The latter is usually obtained through variationally optimizing \textit{non-aufbau} electron configurations via the maximum overlap method.\cite{gilbert_self-consistent_2008,besley_self-consistent-field_2009} However, the application of DFT methods to larger systems is limited by the cubic scaling with respect to the number of electrons. Furthermore, DFT is a single-determinant method, which limits its applicability to strongly correlated\cite{li_multireference_2025,huang_improved_2025} and open shell systems,\cite{fouda_observation_2020,casanova-paez_core-excited_2025,ojha_extended_2025} as well as core-excitations\cite{zamani_assessing_2022,pak_role_2024,nascimento_resonant_2021} with multiconfigurational electronic states.

As ML methods have now established themselves as low cost alternatives to achieving DFT level accuracy that can make predictions at close to no computational cost, they have thus naturally been applied to ``\textit{forward}'' structure-to-spectrum mapping in x-ray spectroscopy.\cite{penfold_machine-learning_2024} Whilst there have been numerous developments in predictive models for x-ray absorption spectroscopy (XAS)\cite{rankine_deep_2020,rankine_accurate_2022,madkhali_role_2020,carbone_machine-learning_2020,kotobi_integrating_2023-1,kharel_omnixas_2025,luder_machine_2025,zhan_graph_2025,gleason_cuxasnet_2025,prange_toward_2025}, machine learning predictions of XPS are less common. Sun \textit{et al.}\ applied gradient boosting predictions of CEBEs to molecular dynamics simulations to predict the XPS of the solid electrolyte interface of lithium metal batteries\cite{sun_machine_2022}. Golze \textit{et al.}\ used kernel ridge regression (KRR) models with smooth overlap of atomic positions (SOAP)\cite{bartok_representing_2013} descriptors to predict CEBEs at the highly accurate \textit{GW} level\cite{mejia-rodriguez_scalable_2021,mejia-rodriguez_basis_2022} for CHO containing disordered materials and molecules, achieving sub 0.1 eV accuracy\cite{golze_accurate_2022}. This model was also integrated with a machine learning potential and molecular dynamics/Monte Carlo simulations to identify which oxygenated amorphous carbon structures were compliant with experimental XPS predictions\cite{zarrouk_experiment-driven_2024}, demonstrating an impressive dual ``\textit{forward}'' and ``\textit{backward}'' mapping functionality for real world applications. KRR has also been successfully applied to molecular CEBE predictions trained with $\Delta$SCF.\cite{tripathy_chemical_2024,porcelli_photoemission_2025}

The application of ML models, such as KRR and deep neural networks (DNNs), to x-ray spectroscopy requires a representation of the local bonding environment around the absorbing atom site for the model's input. These often use atomic descriptors such as the SOAP or the local many-body tensor representation (LMBTR)\cite{huo_unified_2022}, which encode the local atomic environment via expansions of the atomic density or distributions of structural motifs. These descriptors require careful determination of the parameters defining the environment, such as a cutoff radius, and the learned models will be sensitive to deviations in this parameterization, which limits their generalization and transferability. Tripathy \textit{et al.}\cite{tripathy_chemical_2024}\ however, demonstrated that a KRR model trained on local environment descriptors learned by a graph neural network (GNN) could predict CEBEs, suggesting that GNN representations offer an alternative to hand-crafted descriptors.

\begin{figure*}
    \centering
    \includegraphics[width=1.0\textwidth]{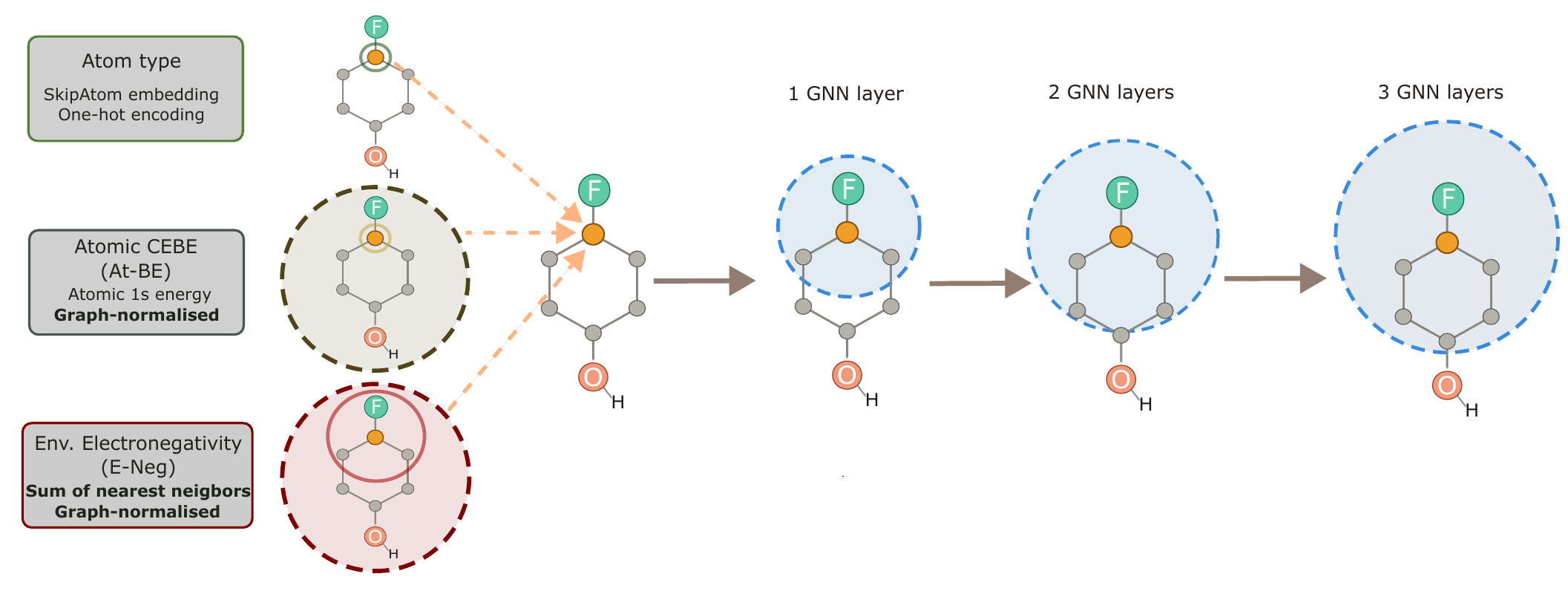}
    \caption{Schematic showing how the graph normalized node features encode molecule specific information at each atom before the number of GNN layers defines the models receptive field.}
    \label{fig:env}
\end{figure*}

GNNs have seen widespread adoption in molecular property\cite{buterez_transfer_2024} and spectroscopy prediction\cite{schutt_equivariant_nodate,xu_pretrained_2025} applications, including XAS.\cite{kotobi_integrating_2023-1,zhan_graph_2025,gleason_cuxasnet_2025} GNNs alleviate the requirement for tuning appropriate descriptors of the local environment by embedding the molecular structure directly into the architecture via the representation of atoms and bonds as the nodes and edges of the network. Moreover, modern equivariant GNN architectures\cite{satorras_en_2022,schutt_equivariant_nodate,batatia_mace_2023,geiger_e3nn_2022-1} incorporate 3D geometric information into the message passing operations in a manner that respects the rotational and translational symmetries of molecular properties, enabling the encoding of higher-order structural effects beyond the molecular graph topology.

Furthermore, the architecture of GNN models inherently relates to the local nature of x-ray spectroscopy observables. In XPS, the effects of the local bond environment are encoded into a single scalar binding energy observable, which primarily depends on the electronic and nuclear structure of the ground state. Jangid \textit{et al.}\cite{jangid_core_2024} have shown that accurate core-ionization energies can be obtained using embedding approaches that apply a more accurate treatment to the local region around the core hole. Similarly, GNN architectures can be tuned to select the essential local environment effects governing the CEBE chemical shifts by naturally encoding the locality of the chemical environment. The number of GNN message passing layers corresponds to the extent of local bond environment considered, as it defines the topological (bond) radius ($r$) of the model's receptive field (see Figure \ref{fig:env}). For example, using a 1 layer GNN each atom receives information from its nearest bonding neighbors ($r=1$) and using a 2 layer GNN each atom receives information from the second nearest bonding neighbors ($r=2$) and so on, though information from more distant neighbors is progressively attenuated through successive aggregation steps due to the over-squashing phenomenon\cite{alon_bottleneck_2021}. This immediate, ``out-the-box'', degree of interpretability makes GNN architectures highly appealing for x-ray spectroscopy applications.

Herein, we train the equivariant GNN (EGNN)\cite{satorras_en_2022} model on multiconfiguration pair-density functional theory (MC-PDFT)\cite{li_manni_multiconfiguration_2014,sharma_multiconfiguration_nodate} carbon 1$s$ CEBEs for 2116 molecules (8637 C atoms). The molecular geometries were taken from the QM9 database of small organic molecules  (H, C, N, O or F containing molecules with maximum 9 heavy atoms).\cite{ramakrishnan_quantum_2014} MC-PDFT is an efficient multireference electronic structure method which computes the energy from the one- and two-particle reduced density matrices, along with optimized orbitals, of a multireference wave function. In the present case, the multireference wave function is a restricted active space self consistent field wave function, which can optimize core-hole states by restricting the number of holes or electrons in an active orbital subspace\cite{josefsson_ab_2012,koulentianos_high_2020,tenorio_multi-reference_2021}. We recently benchmarked this method for its efficiency and experimental accuracy for Auger-electron spectroscopy (AES) simulation.\cite{fouda_computation_2025} CEBEs are included in the computation of AES, and the method gave an experimental carbon 1$s$ CEBE mean absolute error (MAE) of 0.27 eV against 21 molecules. This is promising considering that carbon 1$s$ CEBEs are approximately between 290 and 300 eV, with an experimental uncertainty defined by the 0.1 eV lifetime broadening width of carbon 1$s$ hole states. This method is thus used for the GNN training data in the present study. 

We evaluate the model against 570 experimental CEBEs in 113 molecules with up to 45 atoms, achieving a MAE of 0.33 eV on a held-out test set. This demonstrates the EGNN has good size transferability to larger organic molecules and shows potential for overcoming size scaling limitations suffered by quantum chemical simulations. This size transferability is showcased on the 45 atom avobenzone tautomers, demonstrating the model's ability to rapidly interrogate complex organic molecules where imprecise DFT assignments were previously reported.\cite{abid_electronion_2020} The EGNN is benchmarked against its rotationally invariant analogue (IGNN) and the SOAP-KRR method for data efficiency and stability to non-equilibrium geometries. Whilst all three models show similar training data size dependencies, the SOAP-KRR model is less stable to geometric distortions beyond the equilibrium and is less accurate for the 45 atom enol- and keto- avobenzone molecules. Furthermore, we show that two chemically informed, graph-normalized node features encode molecule specific information at each atom and reduce the EGNN's dependence on the number of message passing layers. By showing that ML models can be efficiently trained on small multireference quantum chemistry datasets and retain their accuracy when applied to larger systems, we lay the foundation for future ML models to make experimentally accurate predictions on highly correlated and open-shell systems.

\section{\label{sec:comp}Computational Details}

\subsection{Dataset}

\begin{figure*}
    \centering
    \includegraphics[width=1.0\textwidth]{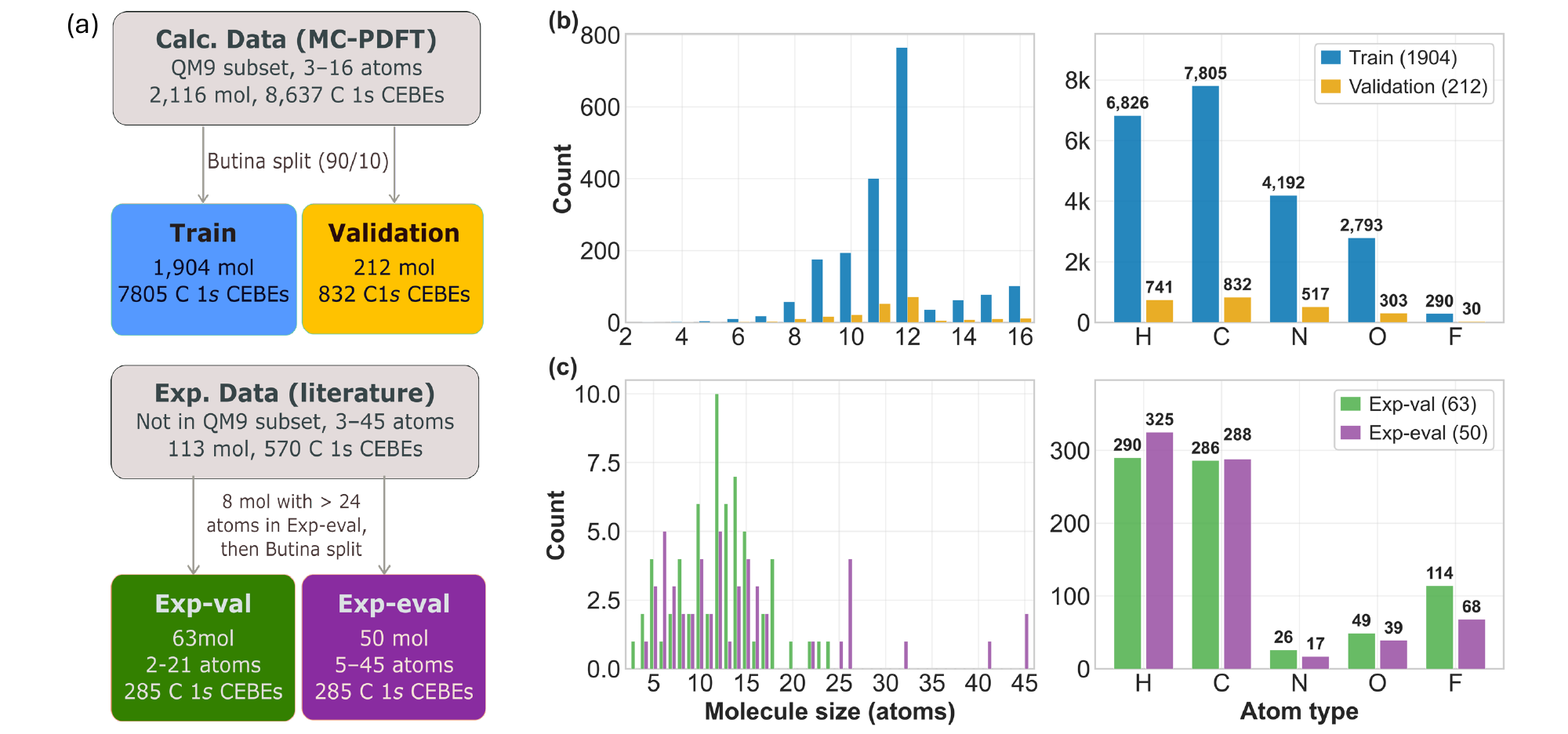}
    \caption{(a) Diagram showing how the calculated data is split into train and validation sets and the experimental data is split into validation (exp-val) and evaluation (exp-eval) data sets. (b) and (c) show bar charts of molecule size (left) and atom type (right) counts of the calculated train (blue) and validation (orange), and experimental validation (Exp-val green) and evaluation (Exp-eval) datasets respectively. The total number of molecules in each set is given in parentheses.}
    \label{fig:data}
\end{figure*}

The training data contains 2116 molecules of 4-16 atoms from the QM9 database.\cite{ramakrishnan_quantum_2014} Only carbons were assigned CEBE, yielding a total of 8637 data samples. The CEBEs were calculated using the MC-PDFT\cite{li_manni_multiconfiguration_2014} implementation in OpenMolcas\cite{li_manni_openmolcas_2023} using the tPBE0 functional and the ANO-RCC-VTZP basis set. Details of the level of theory used are given in a previous study benchmarking MC-PDFT for the computation of AES.\cite{fouda_computation_2025}

Moreover, the present work utilizes the large amount of experimental data available for carbon 1$s$ BEs for gas phase molecules in the literature and corresponds to an additional 570 experimental CEBEs in 113 molecules. This experimental dataset contains molecule sizes of 3-45 atoms; the CEBEs are predominantly from the database compiled by Jolly \textit{et al.},\cite{jolly_core-electron_1984} with additional contributions from more recent literature.\cite{ganguly_coincidence_2022,hitchcock_carbon_1986,bearden_reevaluation_1967,travnikova_esca_2012,naves_de_brito_experimental_1991,vall-llosera_c_2008,abid_electronion_2020} Supplementary Information (SI) Table S1, contains the full set of evaluation molecules and experimental CEBEs with the specific citations. The optimized, gas-phase geometries of the evaluation molecules were obtained by DFT with the PBE0 functional with D3BJ dispersion corrections, the def2-TZVP basis set with the def2/J auxillary basis set and the RIJCOSX approximation in the ORCA software.\cite{neese_orca_2012} We ensure that none of 113 molecules with experimental CEBEs are contained in the sample from the QM9 database used for the calculated training and validation CEBEs.

Figure \ref{fig:data} (a) shows how the calculated and experimental data sets were split in this work, (b) and (c) show the molecule size (by atom number) and atom type counts for the split calculated and experimental data respectively. A 10-fold cross-validation resulted in 1904 calculated training and 212 validation dataset molecules. The splitting used the Butina method, which generates a Morgan fingerprint\cite{rogers_extended-connectivity_2010} for each molecule and clusters the fingerprints with RDKit's Butina clustering\cite{butina_unsupervised_1999} implementation. Molecules in the same cluster are separated into either the train or validation sets to ensure structural diversity between the datasets. The experimental data was split into validation (exp-val) and evaluation (exp-eval) datasets. The exp-val data is used as a secondary metric (alongside the calculated validation dataset loss) for selecting the calculated train and validation data Butina split fold and for determining the node feature specification (discussed below). The exp-eval data is used as a hold-out test set. In order to test the model's transferability to larger molecules, all 8 molecules with more than 24 atoms were placed into exp-eval and the remaining assignments were determined by the Butina splitting method. 50 molecules went into the exp-val set and the remaining 63 went into the exp-eval dataset; both experimental datasets contain 285 carbon $1s$ BEs and show different molecular size distributions but similar atom type distributions. SI Table S1 indicates whether the molecule was included in the exp-val or exp-eval sets. The data for this work is publicly available\cite{fouda_2026_19688196} and is automatically downloaded for training by the AugerNet code for GNN CEBE predictions.\cite{fouda_2026_19689244}

To assist the results discussion, a SMARTS string\cite{daylight_chemical_information_systems_smarts-language_2019} pattern matching algorithm was used to assign a carbon environment (functional group) classification to each carbon atom in the data. SI Table S1 contains the class assignments for the experimental data, which served as a benchmark for the pattern matching algorithm. The algorithm also assigns a priority score to each environment class used when a carbon matches multiple environments; more constrained environments, like C$=$O and aromatic environments, have a higher priority. SI Table S2 contains the 36 environments, the corresponding SMARTS patterns, the priority score, and the counts in the calculated and experimental datasets. The calculated data is dominated by aromatic nitrogen (arom N) and alkyne environments containing 1855 and 1014 out of the total 8637 carbons, respectively. The experimental data is dominated by aromatic carbons, containing 219 out of the total 570 carbons in this dataset.

\subsection{Molecular Graphs}

Molecular graphs for the calculated and experimental datasets were generated from xyz files and three atom type representations were tested in the node feature embedding: the 200- and 30-dimensional pre-trained SkipAtom vectors\cite{antunes_distributed_2022}, which are a learned distributed atom type representation inspired by the natural language processing Skip-gram-style model, and a simple one-hot encoding vector of the 5 atom types (H,C,N,O,F). Furthermore, the inclusion of two additional scalar features to the node embedding was explored: the atomic CEBE (At-BE) and the environment electro-negativity (E-neg). The At-BE feature acts as an additional, physically motivated atom-type descriptor. In this work, we take the ANO-RCC reference orbital energies as atomic binding energies. These values are available in the OpenMolcas~\cite{li_manni_openmolcas_2023,sethio_story_2024} basis-set library and are also tabulated in our AugerNet code. Note that the ANO-RCC basis sets are constructed from atomic natural orbitals obtained for different electronic states at CASPT2 level, with scalar relativistic effects included. Hence, from a Koopmans-theorem perspective, these reference orbital energies may therefore be viewed as approximate atomic binding energies.
The E-neg feature is the bond order (BO) weighted sum of the differences in the Pauling electronegativity\cite{pauling_nature_1932} ($X$) between each atom ($i$) and its nearest bond neighbors ($j\in N$) via $\sum^{N}_{j} = BO(X_{i} - X_{j})$.
Where the BO is determined from the 3D geometry using RDKit in AugerNet.

The inclusion of the E-neg feature arises from the following chemical intuition: a lower atomic electron density reduces the shielding experienced by the core electrons and increases the CEBE, thus an increased neighborhood electronegativity relates to a reduction in the atomic electron density. The E-neg\ feature thus acts as a pre-tabulated approximation to the atom site electron density. The scalar At-BE and E-neg\ features were normalized by the mean and standard deviation of the values within each graph, encoding both atom and molecule specific information to each atom. The edge features are described by a one-hot encoding of 4 bond types: single, double, triple and aromatic.

The output labels are the difference in the atomic (At-BE) and molecular (Mol-BE) CEBEs ($\Delta$CEBE$=$At-BE$-$Mol-BE), which were normalized by the mean and standard deviations from across the whole dataset with values of 15.33 and 1.65 eV respectively. 

\subsection{Model Architecture and Training}

The model architecture is an implementation of the equivariant GNN (EGNN)\cite{satorras_en_2022}, which includes 3D geometry effects via an equivariant coordinate update message operation and circumvents the requirement for spherical harmonic and radial basis expansions present in other equivariant GNN architectures\cite{passaro_reducing_2023,batatia_mace_2023,geiger_e3nn_2022-1}. We refer the reader to the original article for more details on EGNN and provide a summary of our implementation of the architecture in the AugerNet software package\cite{fouda_2026_19689244} herein.

Initially, the node features ($h$) with dimension $\mathbb{R}^{d_{n}}$ are linearly projected to an embedding space of $\mathbb{R}^{d}$ (where $d$ is fixed to 64 in this work). At each message passing layer ($l$), the node embedding for     each atom $h_{i}^{l}$ is augmented with the edge features ($a_{ij}$) and absolute squared difference between atom positions $\big|\big|x_{i}^{l}-x_{j}^{l}\big|\big|^{2}$, then passed as an input through the edge operation ($\phi_{e}$) to form the edge messages ($m_{ij}$) between two nodes via,
\begin{equation}\label{eq:edge}
    m_{ij} = \phi_{e} \Big(h_{i}^{l},h_{j}^{l},\big|\big|x_{i}^{l}-x_{j}^{l}\big|\big|^{2},a_{ij}\Big).
\end{equation}
$\phi_{e}$ involves a two layer multilayer perceptron (MLP), with ReLU activation between the dense layers, and outputs learned messages (with dimension $\mathbb{R}^{d}$) between each node and its nearest neighbors.

Passing the absolute squared bond distance through the edge operation encodes structural information affecting the CEBE into the model whilst respecting the invariance of the CEBE to 3D rotation of the coordinates. Higher-order structural information affecting the CEBE is then encoded into the model by an equivariant coordinate update,
\begin{equation}\label{eq:coord}
x_{i}^{l+1} = x_{i}^{l} + \frac{1}{M_{i}}\sum_{j\neq i}\big(x_{i}^{l}-x_{j}^{l}\big)\phi_{x}(m_{ij}).
\end{equation}
Equation \ref{eq:coord} updates the position of each atom by the sum of the relative differences weighted by $\phi_{x}$ and mean aggregated by normalizing the summation with the total number of bond neighbors for each atom ($M_{i}$). $\phi_{x}$ is the coordinate message operation which takes $m_{ij}$ as input and involves a two layer MLP with ReLU activation, outputting a scalar ($\mathbb{R}^{1}$) passed through a tanh activation to prevent the divergence of updated coordinates during training. 

The edge messages from each atom's neighbors are then aggregated by a summation $m_{i} = \sum_{j\ne i} m_{ij}$, concatenated with the node embedding of the previous layer ($h_{i}^{l}$) and passed to the node update operation ($\phi_{h}$), which involves a two-layer MLP with ReLU activation and output dimension $\mathbb{R}^{d}$. A residual connection adds the node embedding from the previous layer to the MLP output, which is passed through layer normalization,
\begin{equation} 
    h_{i}^{l+1} = \text{LayerNorm}\Big(h_{i}^{l} + \phi_{h}\big(h_{i}^{l}, m_{i}\big)\Big).
\end{equation} 
The updated node embeddings are then passed to a prediction head which linearly projects each node embedding from $\mathbb{R}^{d}$ to a scalar $\Delta$CEBE value for each atom.

\begin{figure}
    \centering
    \includegraphics[width=1.0\columnwidth]{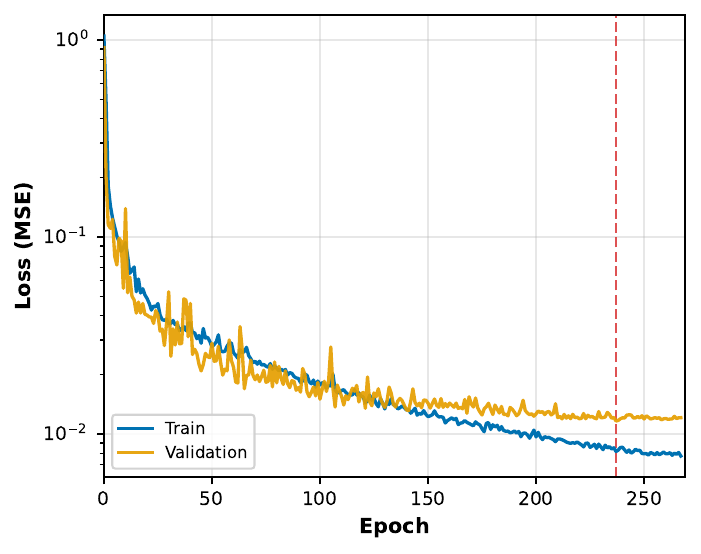}
    \caption{ Train (blue) and validation (orange) loss curves for a 3 layer EGNN model with a (Skipatom-200, At-BE, E-neg) node feature specification. The red dashed line indicate the point at which the 30 epoch early stopping patience selected the model at epoch 237 with train and validation MSE losses of 0.008 and 0.012 respectively.}
    \label{fig:train}
\end{figure}

\begin{figure*}[!htpb]
    \centering
    \includegraphics[width=1.0\textwidth]{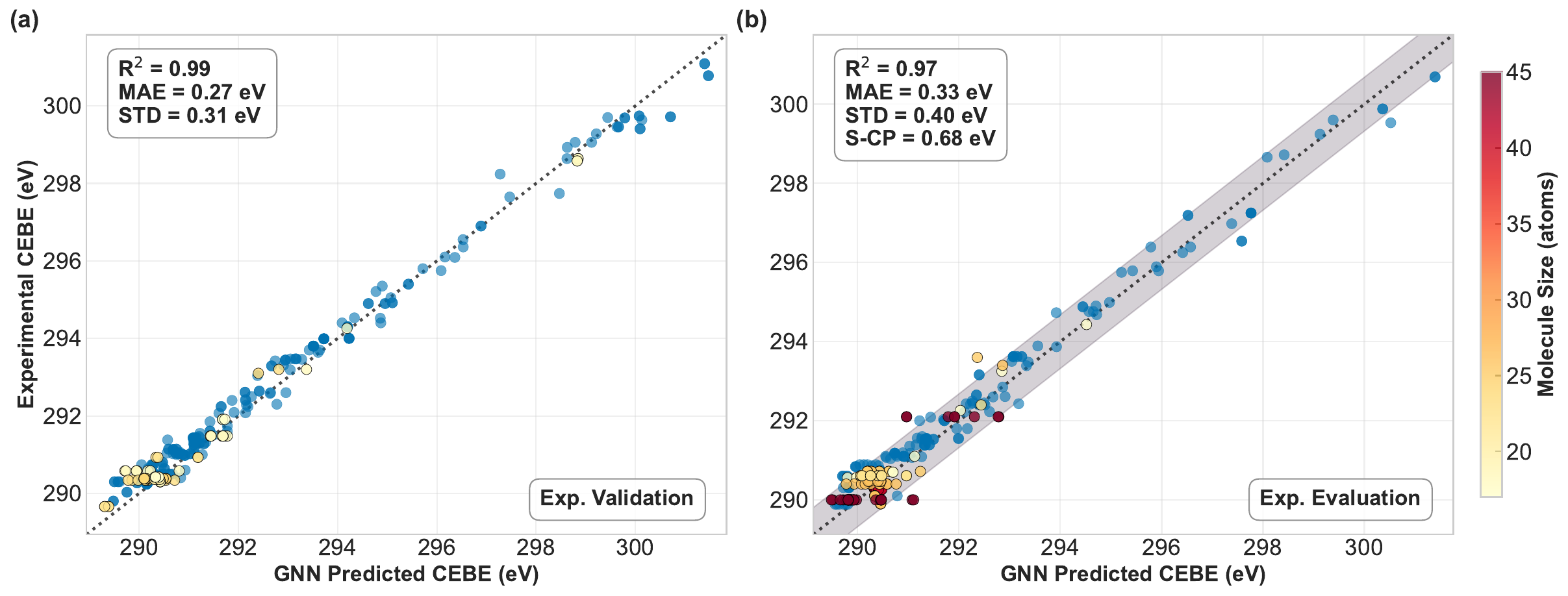}
    \caption{Scatter plot of experimental CEBE's against 3 layer EGNN predictions using a 3 layer model with a (Skipatom-200, At-BE, E-neg) node feature specification for the (a) the experimental validation set and b) the experimental evaluation set. The blue scatter points indicate CEBE values in molecules with up to 16 atoms and the yellow-orange-red color gradient points indicate molecules with more than 16 atoms. The black dotted line is the y$=$x line and the shaded region in (b) indicates the model's quantified uncertainty of $\pm0.68$ eV determined by split conformal prediction (S-CP).}
    \label{fig:scatter}
\end{figure*}

The models were trained with a mean squared error (MSE) loss function of $\mathcal{L} = \big|\big|\Delta CEBE^{True}-\Delta CEBE^{Pred}\big|\big|^{2}$ and ran for a maximum of 300 epochs. Early stopping of the training was applied if the validation loss did not improve within 30 epochs, which improved the training efficiency and prevented overfitting. A dropout rate of 0.1 was applied to the node embeddings between each message passing layer, and gradient norms were clipped to a maximum of 0.5 to stabilize training. The batch size is 24 and the AdamW optimizer\cite{loshchilov_decoupled_2019} with a weight decay of $5\times10^{-4}$ and $\beta$ parameters of $\beta_{1}=0.9$ and $\beta_{2}=0.999$ were used. A cosine annealing learning rate scheduler was applied, which uses a 10 epoch warm-up period with maximum and minimum learning rate of $1\times10^{-3}$ and $1\times10^{-7}$ respectively.

The training of a 3 layer EGNN model on the calculated training data is shown in Figure \ref{fig:train} for the train (blue) and validation (orange) loss curves for this model. The red dashed line indicates the point (epoch 237) at which the 30 epoch early stopping patience selected the model. The separation between the respective train and validation losses of 0.008 and 0.012 is relatively small, indicating that the model has not overfit the training data. 

\subsection{Baseline Regression Model}

The EGNN is bench-marked against the kernel ridge regression (KRR) model with SOAP input descriptors; this method has been used in previous studies on ML CEBE predictions.\cite{golze_accurate_2022,zarrouk_experiment-driven_2024,tripathy_chemical_2024,porcelli_photoemission_2025} Herein we follow the implementation by Porcelli \textit{et al.},\cite{porcelli_photoemission_2025} which uses the SOAP implementation in the DScribe software package,\cite{dscribe,dscribe2} and we refer the reader to these publications for more details. SOAP describes the local bonding environment for each carbon by using an expansion of a gaussian smeared atomic density based on spherical harmonics and radial basis functions.  The descriptors take the following parameters which are determined by hyperparameter optimization: $r_{cut}$ is the cutoff radius (in \AA) for the $n_{max}$ number of radial basis functions, $l_{max}$ is the maximum angular degree used in the spherical harmonics expansion, and $\sigma$ is the width of the Gaussian smoothed atomic density. Our SOAP parameter optimization uses a $r_{cut} = [4.0,6.0,8.0]$, $n_{max} = [6,8,10]$, $l_{max} = [4,6,8]$, and $\sigma = [0.025,0.05,1.0]$ search grid. This was done in combination with the KRR hyperparameter optimization. The scikit-learn\cite{JMLR:v12:pedregosa11a} implementation of the KRR model was used, KRR uses a Gaussian kernel function ($k(p_{i},p_{j})$) to estimate the similarity between the SOAP descriptors ($p$) between all the carbon atoms in the training data,
\begin{equation}
   k(p_{i},p_{j}) = e^{-\gamma||p_{i}-p_{j}||^{2}_{2}},
\end{equation}
where $\gamma$ is the kernel width parameter. This is used to generate a $N^{train}\times N^{train}$ kernel matrix \textbf{K} which trains the model via,
\begin{equation}
   (\textbf{K}+\alpha \textbf{I})\textbf{w} = \textbf{y},
\end{equation}
where \textbf{y} is the vector of CEBEs, \textbf{I} is the $N^{train}\times N^{train}$ identity matrix, \textbf{w} are the regression weights, and $\alpha$ is the regularization parameter used to prevent overfitting. As the EGNN model is trained with a calculated validation set for the model hyperparameter tuning and the exp-val dataset is used for the node feature selection and cross validation data split, we reflect this in the SOAP-KRR parameter search for consistency. The same train and validation data-split as the EGNN was used, the KRR parameters search was performed over a logarithmic grid and assessed against the calculated validation set, and the SOAP parameters were assessed against the exp-val data set. The final SOAP parameters were $r_{cut}=4.0$ \AA, $n_{max}=8$, $l_{max}=6$ and $\sigma=0.025$ and KRR parameters were $\gamma=10.0$ and $\alpha=10^{-11}$, which were tested against the exp-eval dataset. The scripts for the SOAP-KRR implementation can be found at [https://github.com/afouda11/AugerNet].

\section{Results\label{sec:results}}
\noindent

First we show the application of the model trained in Figure \ref{fig:train} to the experimental validation (exp-val) and evaluation (eval) molecules in Figure \ref{fig:scatter} (a) and (b) respectively, which plots the experimental CEBEs against the EGNN predictions. The blue scatter points indicate CEBEs from molecules with up to 16 atoms, the yellow-orange-red color gradient points indicate molecules with more than 16 atoms and the black dotted line is the y$=$x line. The EGNN predictions against both experimental datasets fit to the $x=y$ line extremely well, with an exp-val $R^2$ and MAE values of 0.99 and 0.27 eV, and eval $R^2$ and MAE values of 0.97 and 0.33 eV. The performance on the held-out exp-eval data is impressive considering that training data contains a maximum of 16 atoms per molecule and the model provides predictions with near experimental accuracy on molecules with up to 45 atoms. This demonstrates the potential for GNNs to overcome the size scaling limitations suffered by quantum chemical simulations of x-ray spectroscopy observables. The shaded area around the y$=$x line is the model's quantified uncertainty of $\pm0.68$ eV for experimental carbon 1$s$ binding energies in organic molecules.  This value was determined by split-conformal prediction
(S-CP),\cite{10.1007/3-540-36755-1_29,NEURIPS2019_8fb21ee7} taking the absolute
prediction errors $R_i = |E_i^{\mathrm{exp}} - E_i^{\mathrm{pred}}|$ over the set of 249
exp-eval CEBEs ($N^{exp-eval}$) as a calibration set. The quantified uncertainty is the
$k^{\mathrm{th}}$ smallest value of $R_i$, where $k$ is determined by
\begin{equation}
   k = \lceil (1-\alpha^{CP})(N^{exp-eval} + 1) \rceil.
\end{equation}
Setting $\alpha^{CP}$ to 0.1 gives $k=225$ and an S-CP uncertainty of $\pm0.68$ eV, such that
90$\%$ of the calibration errors lie within this interval. In principle, a prediction on a new organic molecule then has a 90$\%$ probability of being within $\pm0.68$ eV of the experimental value. However, this guarantee holds only if the new molecule is exchangeable with the calibration set, which only includes organic molecules. The interval is also calibrated against experimental values that the procedure treats as exact. These measurements are compiled across several decades and instruments, with reported precisions of $\pm0.02$
to $\pm0.05$ eV\cite{PhysRevA.14.2133,saethre1989effect,myrseth2002adiabatic} and an
intrinsic carbon 1$s$ linewidth of 0.1 eV set by the core-hole
lifetime.\cite{NICOLAS2012267} We note that the EGNN predictions on the 45 atom enol and keto avobenzone tautomers (maroon scatter points in Figure \ref{fig:scatter} (b)) were excluded from $N^{exp-eval}$ in the S-CP analysis, as approximate assignments for either C-O and C=O carbons at 292.1 eV and alkyl and benzene ring sites at 290.0 eV were given previously\cite{abid_electronion_2020}. Later in the text, we use the EGNN to interrogate the precise assignments of these systems. The full set of exp-val and eval EGNN predictions, alongside molecule names and carbon environment classes is given in SI Table S1.

\begin{figure}[!htbp]
    \centering
    \includegraphics[width=1.0\columnwidth]{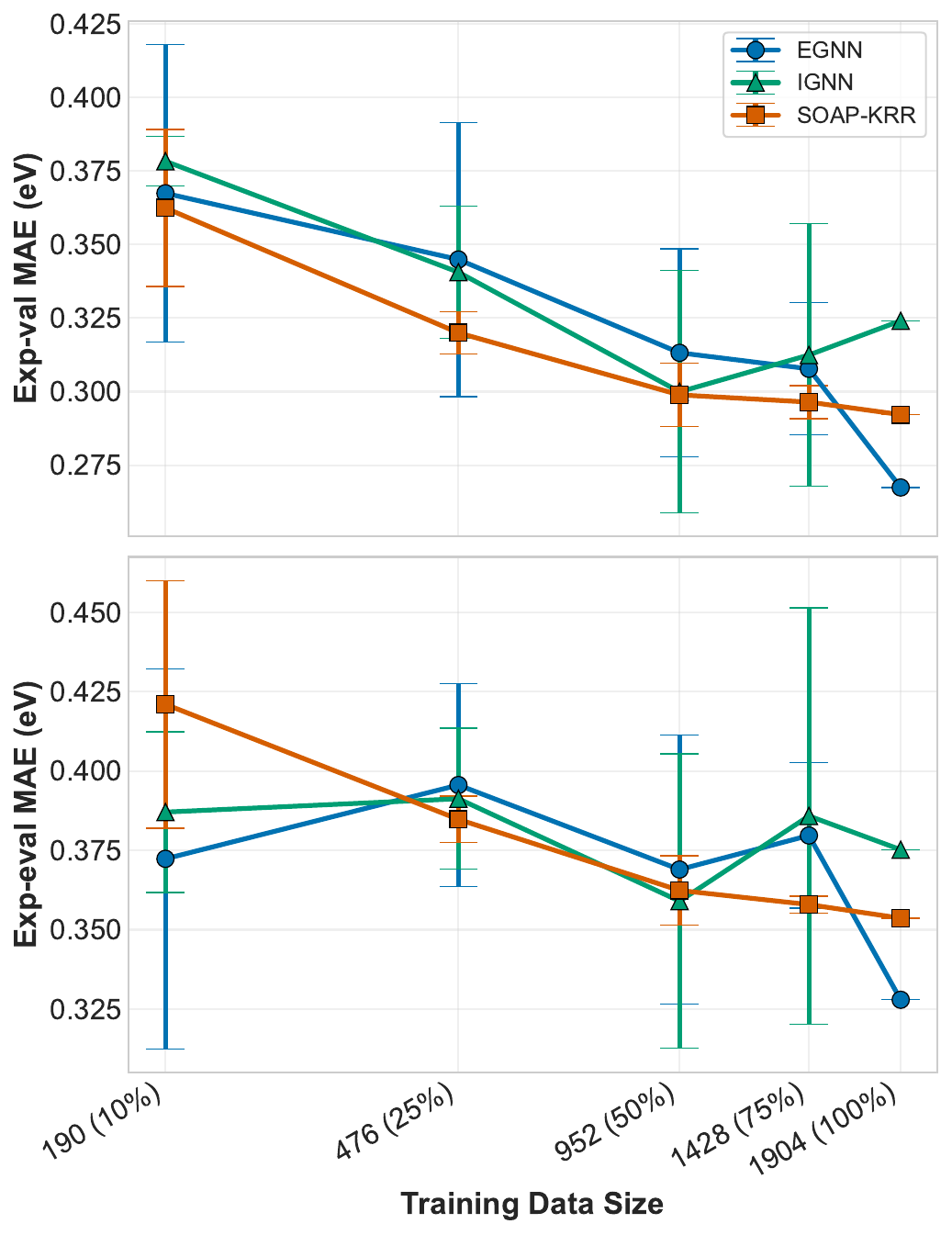}
    \caption{Training data efficiency plots for the EGNN, IGNN and SOAP-KRR models for the exp-val (top) and exp-eval (bottom) datasets. The less than 100$\%$ data points were determined by randomly sampling a subset of molecules and taking a 4 seed average. The error bars are the standard deviation of this averaged sampling.}
    \label{fig:data_eff}
\end{figure}

The data efficiency of the EGNN model is given in Figure \ref{fig:data_eff}. The model is compared to two other models: its invariant analogue (IGNN), which has the same architecture minus the coordinate update steps, thus only including the constant absolute squared difference of atom positions, and the SOAP-KRR model. The results in Figure \ref{fig:data_eff} were determined by randomly extracting a set of molecules from the calculated training data and performing a 4-seed average, the standard deviation of this averaging is given by the error bars. All three models give a similar performance and the SOAP-KRR model outperforms the EGNN and IGNN on the Exp-val molecules at 75\% of the training data. Both the EGNN and IGNN models' Exp-val performance varies little with respect to the training data and the small fluctuations in the performance result from the 30 epoch patience used to stop the model training. Overall all three models show excellent data efficiency for predicting carbon 1$s$ binding energies.

\begin{figure*}
    \centering
    \includegraphics[width=\textwidth]{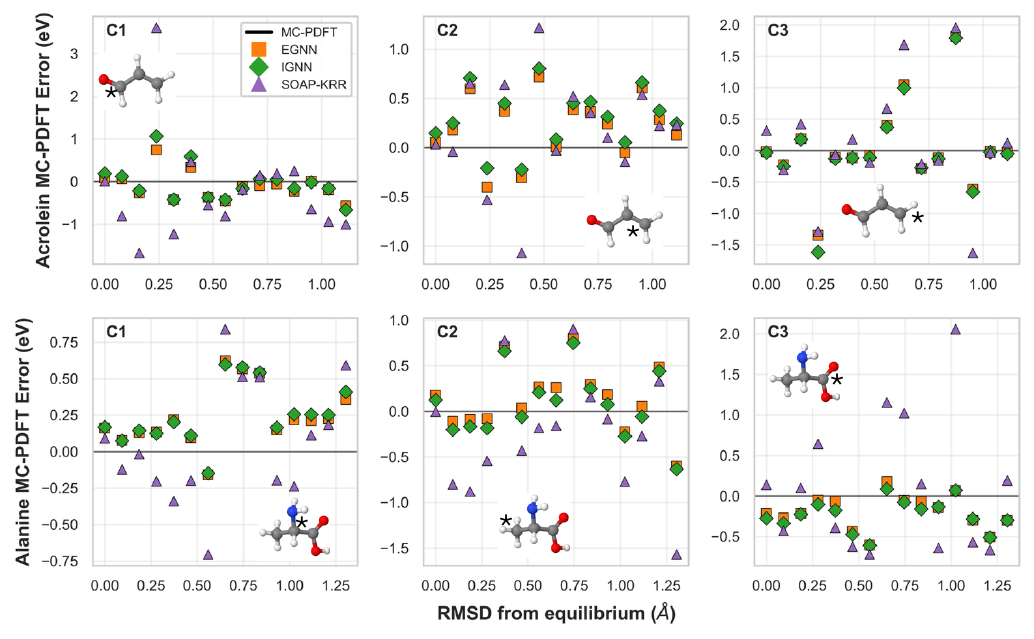}
    \caption{Comparison of the EGNN (orange), its invariant analogue (IGNN) (green), and the SOAP-KRR model on the non-equilibrium geometries of the acrolein and alanine. The predicted errors are with respect to the MC-PDFT tPBE0 training data level of theory (solid black line). The star next to each carbon indicates which carbon each panel corresponds to. The non-equilibrium geometries were taken from the WS22 dataset.}
    \label{fig:pes}
\end{figure*}

The transferability of the EGNN, IGNN and SOAP-KRR models to non-equilibrium geometries is given for the acrolein and alanine molecules in Figure \ref{fig:pes}. 1 equilibrium and 14 non-equilibrium geometries for each molecule were extracted from the WS22 dataset\cite{PinheiroJr2023}. The errors with respect to the MC-PDFT tPBE0 training data level of theory are plotted against the root mean square deviation (RMSD) for each geometry from the equilibrium geometry. Whilst the EGNN and IGNN models are in close agreement in all cases the SOAP-KRR model results show more significant deviations from the graph models and the calculations. However, in many cases the SOAP-KRR model shows good agreement to the EGNN and IGNN models and calculations, demonstrating good transferability for a descriptor based ML method. The good agreement between the EGNN and IGNN models indicates that only including the structure as fixed bond-lengths sufficiently learns the structure to CEBE relationship, with respect to including E(3)-equivariance in the EGNN model.

% ──────────────────────────────────────────────────────────────────────────────
%  SI Table: Node-feature ablation study
% ──────────────────────────────────────────────────────────────────────────────
\begin{table*}[htbp]
\centering
\caption{Effect of node-feature specifications on a 3 layer EGNN model.
The atom-type column indicates the atom-type representation;
At-BE is the graph-normalized atomic 1s binding energy ($\in \mathbb{R}^{1}$)
and E-neg is the graph-normalized environment electronegativity
($\in \mathbb{R}^{1}$).
$d_x$ denotes the total node-feature dimension.
Val $\mathcal{L}$ is the best MSE reached with patience of 30 epochs and the calculated validation dataset;
exp-val MAE is the mean absolute error on the 50 experimental validation molecules in eV.
The $\Delta$ columns give the relative difference of the val $\mathcal{L}$
and exp-val MAE metrics.}
\label{tab:nodefeat}
\small
\begin{tabular*}{\textwidth}{@{\extracolsep{\fill}}llcccccc@{}}
\toprule
Atom Type & At-BE & E-neg & $d_n$ & Val $\mathcal{L}$ & $\Delta \mathcal{L}$ & Exp-val MAE (eV) & $\Delta $ MAE (eV) \\
\midrule
%% --- Skip-200 group ---
Skip-200 & \checkmark & \checkmark & 202 & 0.012 & 0 & 0.267 & 0.036 \\[2pt]
Skip-200 & \checkmark &  & 201 & 0.014 & 0.002 & 0.264 & 0.033 \\[2pt]
Skip-200 &  & \checkmark & 201 & 0.017 & 0.005 & 0.261 & 0.029 \\[2pt]
Skip-200 &  &  & 200 & 0.028 & 0.017 & 0.306 & 0.074 \\
\midrule
%% --- Skip-30 group ---
Skip-30 & \checkmark & \checkmark &  32 & 0.012 & 0 & 0.275 & 0.044 \\[2pt]
Skip-30 & \checkmark &  &  31 & 0.012 & 0.001 & 0.263 & 0.031 \\[2pt]
Skip-30 &  & \checkmark &  31 & 0.019 & 0.008 & 0.232 & 0 \\[2pt]
Skip-30 &  &  &  30 & 0.030 & 0.019 & 0.314 & 0.082 \\
\midrule
%% --- One-hot group ---
One-hot & \checkmark & \checkmark &   7 & 0.012 & 0.001 & 0.315 & 0.084 \\[2pt]
One-hot & \checkmark &  &   6 & 0.014 & 0.002 & 0.331 & 0.099 \\[2pt]
One-hot &  & \checkmark &   6 & 0.021 & 0.010 & 0.255 & 0.024 \\[2pt]
One-hot &  &  &   5 & 0.029 & 0.018 & 0.300 & 0.068 \\
\midrule
%% --- None group ---
At-BE & \checkmark & \checkmark &   2 & 0.016 & 0.004 & 0.270 & 0.038 \\[2pt]
At-BE & \checkmark &  &   1 & 0.014 & 0.003 & 0.330 & 0.098 \\
\bottomrule
\end{tabular*}
\end{table*}

Unless stated otherwise, a (Skipatom-200, At-BE, E-neg) node feature specification was used for GNN results in this work. This specification was determined from a systematic benchmark on how the node features affect the model performance shown in Table \ref{tab:nodefeat}. The absolute calculated validation loss ($\mathcal{L}$), experimental validation (exp-val) MAE (eV), and their relative differences ($\Delta\mathcal{L}$ and $\Delta$MAE (eV)) are provided. The atom-type representation is given in the first column and inclusion of the At-BE and E-neg scalars is indicated by the check-marks in the second and third columns, respectively. The fourth column has the total node feature dimension $\mathbb{R}^{d_{n}}$. The (Skipatom-200, At-BE, E-neg) node feature specification (top row) gives the lowest calculated validation loss and a good experimental validation MAE. The (Skipatom-30, At-BE, E-neg) model gives an equally low calculated validation loss, but the (Skipatom-200, At-BE, E-neg) model was chosen for the lower experimental validation MAE. The increased size of the Skipatom-200 vector does not result in prohibitively expensive training costs on the small dataset in this study (2116 molecules). When At-BE is used for the atom type representation (bottom two rows), the model has a relatively good performance and is improved when E-neg is included. Table \ref{tab:nodefeat} validates the use of such chemically informed node features, as concatenating the atom type representation with the At-BE and E-neg scalars generally improves the model's performance (with the exception of the increased experimental validation MAE when the One-hot atom type is concatenated with (At-BE) or (At-BE, E-neg)).

\begin{figure}
    \centering
    \includegraphics[width=\columnwidth]{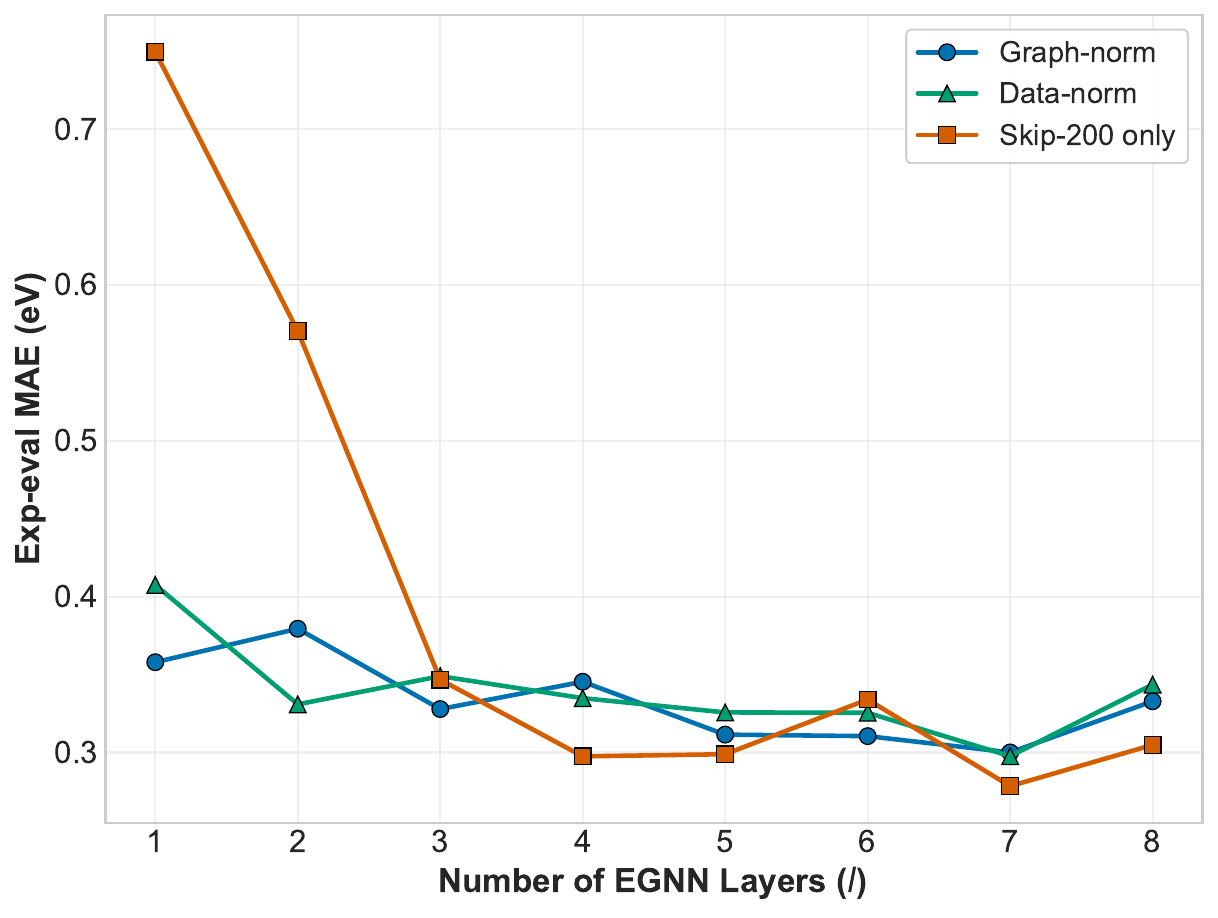}
    \caption{Effect of the number of message passing EGNN layers ($l$) on the MAE of the exp-val set for the (Skipatom-200, At-BE, E-neg) node feature with graph normalized (blue) and data normalized (green) At-BE, E-neg features, and just the Skipatom-200 node feature (orange).}
    \label{fig:scan}
\end{figure}

The E-neg feature involves a summation of the Pauling electro-negativity differences of each atom's nearest neighbors and both scalar features are mean and standard deviation normalized per graph, they therefore encode molecule specific information at each atom. Graph-normalization has been previously shown to be an effective strategy for accounting for the global structure,\cite{chen_learning_2020} which has been applied to molecular property predictions.\cite{fang_geometry-enhanced_2022} Figure \ref{fig:scan} shows the effect of varying the number of EGNN layers on the Exp-val MAE for the (Skipatom-200, At-BE, E-neg) node feature set with graph-normalized At-BE, E-neg features (blue) and data-normalized At-BE, E-neg features (green) and compares them to an model trained with just the Skipatom-200 node feature (orange). The latter is agnostic to the molecular structure beyond each carbon atom, so the receptive field is solely determined by the number of EGNN layers. The (Skipatom-200) feature model requires 3 layers to converge the exp-val and exp-eval MAEs, whilst the graph and data normalized (Skipatom-200, At-BE, E-neg) models show relatively minor performance discrepancies between 1 and 8 layers ($1\leq l\leq 8$). This shows that either encoding nearest neighbor (data-normalized) or molecular level (graph-normalized) information at each atom alleviates the model's dependence on the number of EGNN layers and can enable more efficient training of models on much larger datasets in future studies.

\begin{figure*}[!htbp]
    \centering
    \includegraphics[width=1.0\textwidth]{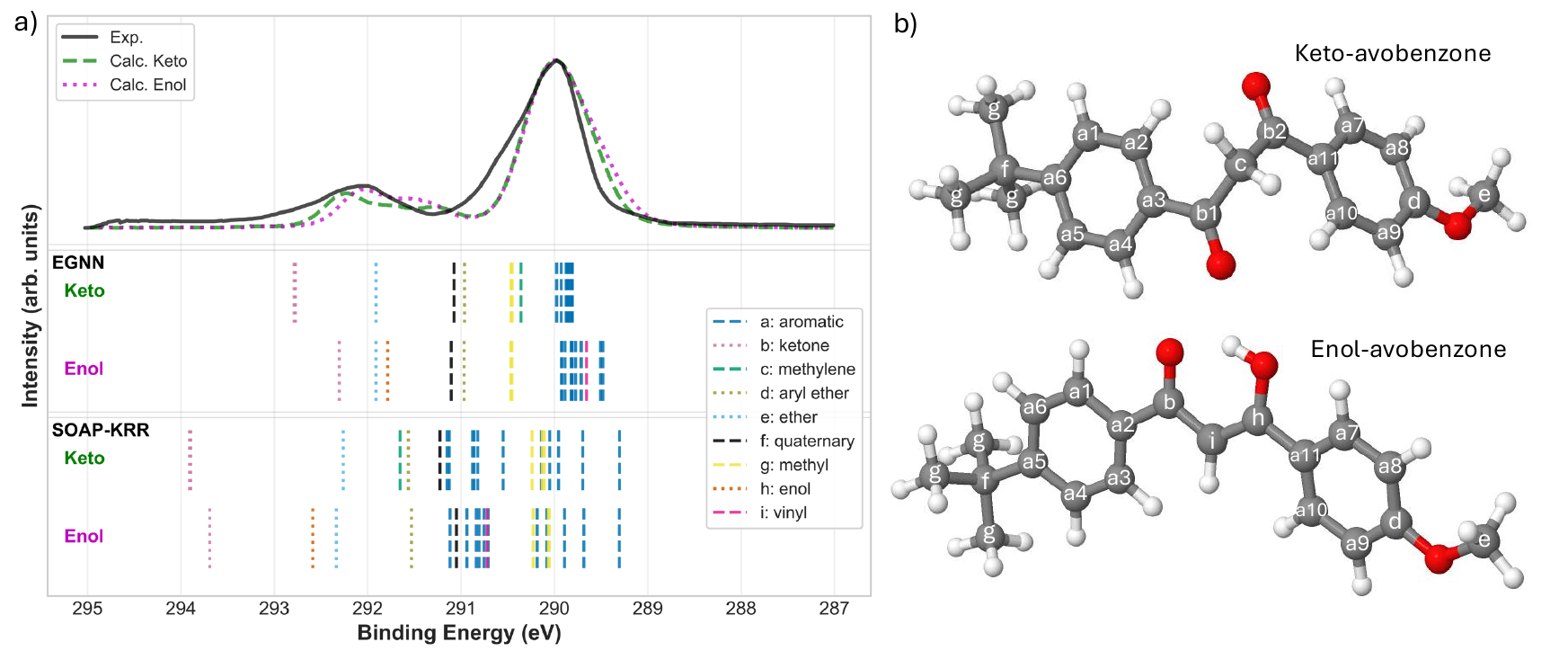}
    \caption{a) Analysis of EGNN predicted CEBE for keto-avobenzone and enol-avobenzone with respect to experimental (top panel: black (both tautomers)) and MP2 calculated (top panel: green dashed (keto-avobenzone), magenta dotted (enol-avobenzone)) carbon 1$s$ XPS spectra\cite{abid_electronion_2020}. The EGNN predictions are given by the dashed colored sticks in the lower panel (upper row: keto-avobenzone, lower row: enol-avobenzone) and are colored by the environment class, determined by the SMARTS pattern matching procedure given in the SI. Labels a-i refer to the specific atom assignments in panel b) to the right and are used for the energy assignments in Table \ref{tab:avo}.}
    \label{fig:avo}
\end{figure*}

We now inspect the performance of the $l=3$ EGNN with the (Skipatom-200, At-BE, E-neg) node features and the SOAP-KRR models to the largest molecules included in the exp-eval dataset, the 45 atom enol-and keto-avobenzone tautomers. In Figure \ref{fig:avo}, the EGNN and SOAP-KRR predictions are considered with respect to previously reported experimental and theoretical spectra.\cite{abid_electronion_2020} (a) shows the experimental XPS spectra (black) containing both tautomers, the Møller–Plesset 2nd order perturbation theory (MP2) calculated keto-avobenzone (green dashed) and enol-avobenzone (magenta dotted) spectra from the previous study. The EGNN and SOAP-KRR predictions are given by the colored dashed and dotted sticks in the lower panel. The sticks are colored by the carbon environment classifications from the SMARTS pattern matching procedure; the a-i labels map the sticks to the carbon atoms labels in (b). Table \ref{tab:avo} provides the CEBE assignments for the specific carbon atoms within each environment. In the previous study, the MP2 calculations provided two approximate experimental CEBE assignments for the two major experimental peaks at 292.1 eV and 290.0 eV. The C-O and C=O sites (b: ketone, d: aryl ether, e: ether, h: enol) were assigned to the 292.1 eV peak and the alkyl and benzene ring sites (a: aromatic, c: methylene, f: quaternary, g: methyl, i: vinyl) were assigned to the 290.0 eV peak, hence the approximate experimental energies and errors in Table \ref{tab:avo}.

\begin{table*}[!htbp]
\centering
\caption{EGNN predicted and approximate experimental carbon 1$s$ CEBEs of keto-avobenzone and enol-avobenzone\cite{abid_electronion_2020}, all energies are given in (eV). The C column has the label for each carbon atom given in Figure \ref{fig:avo} b) and the Env.\ column contains the carbon environment class, determined by the SMARTS patterns in the SI. The errors with respect to the approximate experimental assignments are given in brackets.}
\label{tab:avo}
\footnotesize
\setlength{\tabcolsep}{9pt}
\begin{tabular}{llcrr@{\hskip 2em}llcrr}
\hline
\multicolumn{5}{c}{Keto-avobenzone} & \multicolumn{5}{c}{Enol-avobenzone} \\
C & Env. & $\approx$Exp.\ & EGNN (err) & SOAP-KRR (err) & C & Env. & $\approx$Exp.\ & EGNN (err) & SOAP-KRR (err) \\
\hline
a1   & aromatic     &  290.0 & 289.8 ($-$0.2)     & 289.3 ($-$0.7)     & a1   & aromatic     &  290.0 & 289.8 ($-$0.2)     & 290.7 (0.7)        \\
a2   &              &        & 289.9 ($-$0.1)     & 290.8 (0.8)        & a2   &              &        & 289.9 ($-$0.1)     & 290.8 (0.8)        \\
a3   &              &        & 290.0 ($-$0.0)     & 290.9 (0.9)        & a3   &              &        & 289.8 ($-$0.2)     & 290.8 (0.8)        \\
a4   &              &        & 289.9 ($-$0.1)     & 290.5 (0.5)        & a4   &              &        & 289.8 ($-$0.2)     & 289.3 ($-$0.7)     \\
a5   &              &        & 289.8 ($-$0.2)     & 289.7 ($-$0.3)     & a5   &              &        & 289.9 ($-$0.1)     & 289.9 ($-$0.1)     \\
a6   &              &        & 289.9 ($-$0.1)     & 290.0 ($-$0.0)     & a6   &              &        & 289.8 ($-$0.2)     & 289.7 ($-$0.3)     \\
a7   &              &        & 289.8 ($-$0.2)     & 291.1 (1.1)        & a7   &              &        & 289.5 ($-$0.5)     & 290.9 (0.9)        \\
a8   &              &        & 289.8 ($-$0.2)     & 290.1 (0.1)        & a8   &              &        & 289.7 ($-$0.3)     & 290.2 (0.2)        \\
a9   &              &        & 289.9 ($-$0.1)     & 290.1 (0.1)        & a9   &              &        & 289.8 ($-$0.2)     & 290.1 (0.1)        \\
a10  &              &        & 289.8 ($-$0.2)     & 291.1 (1.1)        & a10  &              &        & 289.5 ($-$0.5)     & 291.1 (1.1)        \\
a11  &              &        & 289.9 ($-$0.1)     & 290.9 (0.9)        & a11  &              &        & 289.9 ($-$0.1)     & 290.8 (0.8)        \\
\cmidrule(r){1-5} \cmidrule{6-10}
b1   & ketone       &  292.1 & 292.8 (0.7)        & 293.9 (1.8)        & b    & ketone       &  292.1 & 292.3 (0.2)        & 293.7 (1.6)        \\
\cmidrule{6-10}
b2   &              &        & 292.8 (0.7)        & 293.9 (1.8)        & d    & aryl ether   &  292.1 & 291.0 ($-$1.1)     & 291.5 ($-$0.6)     \\
\cmidrule(r){1-5} \cmidrule{6-10}
c    & methylene    &  290.0 & 290.4 (0.4)        & 291.7 (1.7)        & e    & ether        &  292.1 & 291.9 ($-$0.2)     & 292.3 (0.2)        \\
\cmidrule(r){1-5} \cmidrule{6-10}
d    & aryl ether   &  292.1 & 291.0 ($-$1.1)     & 291.6 ($-$0.5)     & f    & quaternary   &  290.0 & 291.1 (1.1)        & 291.0 (1.0)        \\
\cmidrule(r){1-5} \cmidrule{6-10}
e    & ether        &  292.1 & 291.9 ($-$0.2)     & 292.3 (0.2)        & h    & enol         &  292.1 & 291.8 ($-$0.3)     & 292.6 (0.5)        \\
\cmidrule(r){1-5} \cmidrule{6-10}
f    & quaternary   &  290.0 & 291.1 (1.1)        & 291.2 (1.2)        & g    & methyl       &  290.0 & 290.5 (0.5)        & 290.1 (0.1)        \\
\cmidrule(r){1-5} \cmidrule{6-10}
g    & methyl       &  290.0 & 290.5 (0.5)        & 290.2 (0.2)        & i    & vinyl        &  290.0 & 289.7 ($-$0.3)     & 290.7 (0.7)        \\
\hline
\end{tabular}
\end{table*}

The EGNN predictions in Figure \ref{fig:avo} (a) show good qualitative agreement to the experimental and MP2 calculated XPS spectra. Two subtle line shape differences between the calculated keto-avobenzone and enol-avobenzone spectra are also reflected by the EGNN predictions. The first being the slope of the calculated keto-avobenzone spectra on the high CEBE side of the 292.1 eV peak is shifted to higher energy than the enol-avobenzone spectra, this agrees with the shift of the EGNN predicted ketone CEBEs (pink dotted lines) for keto-avobenzone to higher energies than the enol-avobenzone ketone CEBE prediction. The second being the slope of the low CEBE side of the 290.0 eV peak is shifted to a lower energy for the calculated enol-avobenzone spectra, which agrees with the aromatic carbon EGNN predictions at lower energies than the keto-avobenzone aromatic carbon predictions. These small shifts are not reflected by the SOAP-KRR results, which show nearly identical CEBEs for the aromatic carbons under the 290.0 eV peak. For the ketone CEBEs, SOAP-KRR predicts both tautomers at significantly higher CEBEs around 294 eV where no intensity for the experimental spectra is observed.

There are two EGNN predictions in both tautomers with approximate errors above 1 eV, which belong to the single aryl ether (d) and quaternary (f) carbons in both molecules. The previous MP2 calculation assigned the quaternary carbons to 290.0 eV peak and the aryl ether carbons to the 292.1 eV peak. The EGNN predicts quaternary CEBEs of 291.33 eV and 291.42 eV for keto-avobenzone and enol-avobenzone respectively (black dashed lines) and aryl ether CEBEs of 290.96 eV and 290.97 eV for keto-avobenzone and enol-avobenzone respectively (olive dotted lines). The EGNN predictions for both these carbon environments lie in the region between the two major peaks. The SOAP-KRR model gives similar predictions to the EGNN for the quaternary carbons and predicts the aryl-ether carbons closer to the 292.1 eV peak that was assigned to these sites in the previous study.\cite{abid_electronion_2020}

To determine whether these erroneous EGNN predictions are systematic for these carbon environments or molecule specific, they can be compared to other examples in the calculated and experimental (SI Table S1) datasets. The single other quaternary carbon in the experimental data is in trimethylacetonitrile (C$_{5}$H$_{9}$N), which has an experimental CEBE of 291.80 eV and a GNN prediction of 291.95 eV (0.15 eV error). The training data also contains 13 quaternary carbons with a calculated CEBE range of 292.60-297.23 eV. However, these quaternary carbons all contain non-H heteroatoms within the $r=3$ local environment and the keto-avobenzone and enol-avobenzone $r=3$ local environments only contain C and H atoms. This potentially highlights a gap in the model's training data for another case where the ``beyond nearest neighbors'' bond environment significantly affects the chemical shift. It is also possible that the large errors for keto-avobenzone and enol-avobenzone aryl ether carbons will arise from a combination of carbon environment gaps in the training data, ``beyond nearest neighbor'' environment effects on the chemical shifts and the approximate assignments of the experimental values for this molecule. This case study demonstrates how EGNN models with demonstrable experimental accuracy can go beyond ``\textit{proof-of-principle}'' demonstrations to rapidly interrogate spectroscopic signals, and that their natural interpretability with respect to local environment effects in x-ray spectroscopy observables offers a simple and intuitive diagnosis of erroneous results and training data gaps.

\section{Conclusion}

We have demonstrated that the EGNN model trained on a small dataset of 2116 molecules with 4-16 atoms, can predict carbon 1$s$ CEBEs with good experimental accuracy on large organic molecules with up to 45 atoms. The MC-PDFT tPBE0 training data level of theory previously showed a 0.27 eV MAE with respect to experiment\cite{fouda_computation_2025} and the model's experimental evaluation MAE on a held-out set of 63 molecules was 0.33 eV, effectively learning the structure-to-binding energy relationship and demonstrating good size transferability. The model was compared to two other models, its rotationally invariant analogue (IGNN) and the SOAP-KRR model for its data efficiency and transferability to non-equilibrium geometries. All three models demonstrate similar data efficiency and the IGNN and EGNN models show better accuracy for the non-equilibrium geometries of the acrolein and alanine molecules. The accuracy of the IGNN model with respect to the EGNN model demonstrates that bond lengths alone are sufficient for the model to learn the structure-to-CEBE relationship captured by an E(3)-equivariant model. The comparable accuracy and transferability of the SOAP-KRR model demonstrate the robustness of descriptor-based models.

%\textcolor{red}{\sout{GNN architectures are advantageous for x-ray spectroscopy as the number of message passing layers can define the topological radius of the model's receptive field, providing an inherent and interpretable connection between the model architecture and the locality of environment effects on the predicted observable. Our layer scan analysis showed how two chemically informed node features: the atomic binding energy and environment electronegativity, encode non-local molecular information when normalized across the graph and enabled the model to distinguish the non-local environment effects with fewer message passing layers, thereby improving its size transferability. This simple feature design encodes higher-order information into the model's receptive field without increasing the number of GNN layers and offers a simple approach to encoding complexity into GNN model predictions of local atomic properties. The layer scan also showed that a model without these chemically informed graph-normalized node features requires at least 3 message passing layers to achieve good experimental accuracy, indicating that CEBE chemical shifts can be influenced by beyond the nearest bonding neighbors environment affects. This is showcased by the aryl fluoride carbons in para-di-substituted fluorobenzenes, where a 1.22 eV range in experimental CEBEs arises from substituents with a topological (bond) radius of 4, from the aryl-fluoride carbon.}}

These results demonstrate that GNN models trained on compact, high-quality MC-PDFT datasets can provide experimentally accurate CEBE predictions while retaining transferability to larger organic molecules. The accuracy and transferability observed here also underscore a broader point: for properties governed by open-shell, core-ionized, or strongly correlated electronic structure, GNNs should be trained on multireference data rather than DFT labels whenever possible. This is an emerging and rapidly expanding area in which machine learning can extend, rather than replace, high-level electronic-structure theory. This strategy opens a route toward fast, chemically interpretable, and quantum-chemically reliable models for X-ray spectroscopy and related electronic-structure observables.

Future studies will involve expanding the training data to non-equilibrium structures which could enable future models in the application of time-resolved XPS studies of ultrafast molecule dynamics.\cite{neppl_time-resolved_2015,gessner_monitoring_2016,rivas_unraveling_2026,freixas_time-resolved_2026} Furthermore, the present work only considers molecular carbon 1$s$ CEBEs from high-level calculations, hence combining this data with available DFT nitrogen 1$s$, oxygen 1$s$ and fluorine 1$s$ molecular CEBE datasets,\cite{porcelli_photoemission_2025,tripathy_chemical_2024} material CEBE datasets\cite{golze_accurate_2022} and the availability experimental molecular CEBEs across the periodic table,\cite{jolly_core-electron_1984} presents an opportunity for EGNN finetuning in the mixed-fidelity data setting for more universal CEBE prediction models.

\begin{acknowledgments}
A.E.A.F. is grateful for the support from the Eric and Wendy Schmidt AI in Science Postdoctoral Fellowship, a Schmidt Futures Program. This work was primarily supported by the U.S. Department of Energy, Office of Basic Energy Sciences, Division of Chemical Sciences, Geosciences, and Biosciences at Argonne National Laboratory, through under contract DE-AC02-06CH11357. This work, at the University of Chicago, was partially supported by the Computational Chemical Sciences Program under Award DE-SC0023382, funded by the U.S. Department of Energy, Office of Basic Energy Sciences, Chemical Sciences, Geosciences, and Biosciences Division.
\end{acknowledgments}

\section*{Data and Software Availability Statement}

The data and the software are freely available at https://zenodo.org/records/19688196  and https://zenodo.org/records/19689244 respectively. The Git repository for the software can be found at https://github.com/afouda11/AugerNet.

\section*{Supplementary Material}

The Supplementary Material (SM) for this work includes a table of the all the experimental binding energies, with the relevant citations and corresponding EGNN predictions for the full experimental dataset used in this work. The SM also contains a table of the SMARTS patterns for each carbon bond environment label and the associated priority scores used in the pattern matching algorithm. 

%\nocite{*}
\section*{References}
\bibliography{literature}% Produces the bibliography via BibTeX.

\end{document}